# R-Matrix Theory for Electron-Ion Collisions in Plasmas

Chao Wu,[1] Wen Hao Xia,[2] Yong Wu,[2,3] Jun Yan,[2,3] Ming Li,[1] Jian Guo Wang,[2] and Xiang Gao[2, *]

1. College of Science, Xi'an University of Science and Technology, Xi'an 710054, China
2. National Key Laboratory of Computational Physics, Institute of Applied Physics and Computational Mathematics, P. O. Box 8009, Beijing 100088, China
3. HEDPS, Center for Applied Physics and Technology, Peking University, Beijing 100084, China

Electron-atom collisions in warm dense plasmas are crucial for astrophysics and controlled fusion research, where calculating short-range scattering matrices under screening plasma potentials is essential. While electron-neutral atom collisions are tractable using the standard Riccati-Bessel wavefunctions in the asymptotic region, electron-ion collisions face challenges due to the extended range of the screened Coulomb potential, which lacks analytical solutions or numerical code packages for asymptotic regular and irregular wavefunctions. We introduce an R-matrix theoretical framework for general screened potentials and develop a numerical method to compute these asymptotic wavefunctions efficiently. Our approach yields short-range scattering phase shifts that remain invariant with respect to the matching point in the asymptotic region. Applying the Debye screening potential as an illustrative example, we calculate elastic and electron-impact excitation collision strengths for H-like ions (He, C, Ne) across varying temperatures and densities. The calculations show that Debye screening systematically modifies resonance structures and progressively lowers excitation thresholds. Nevertheless, the effective collision strengths and rate coefficients exhibit approximate scaling laws. These findings enable convenient access to electron collision data in plasma environments, advancing plasma diagnostics and modeling.

## I. INTRODUCTION

Understanding the detailed dynamics of electron-ion/atom interactions is crucial for diverse plasma applications in fields such as astrophysics [1], fusion research [2], and semiconductor lithography [3]. For such plasmas, whether of astrophysical origin or man-made, it is necessary to understand atomic energy levels and associated collision processes in order to numerically simulate the temporal-spatial motions of plasmas and to diagnose and analyze their conditions.

In practical applications, unlike in isolated systems, the influence of plasma environments on electron collision processes must be considered. For instance, Stark broadening, observed in laboratory and astrophysical plasmas [4], arises from interactions between emitters and nearby ions and electrons. Analyzing line widths in detail enables diagnostics of plasma density and temperature [5]. For rapidly moving electrons, the abrupt phase changes caused by a series of collisions between the electrons and the emitter are the primary cause of spectral line broadening [6]. Baranger [7] and Seaton [8] derived the expression for this effective interaction and emphasized its relationship with the elastic scattering amplitude. Thus, accurate electron-ion/atom collision data in plasmas are crucial for reliable spectral line broadening calculations.

However, plasmas are inherently complex systems, composed of a large number of free electrons, ions, and neutral atoms, which pose a significant challenge in modeling the environmental effects. According to Pines's theory [9], electron behavior in plasmas can be divided into two components based on their density fluctuations. One component is associated with the organized oscillation of the system as a whole, the so-called plasma oscillation. The other is associated with the random thermal motion of the individual electrons, which shows no collective behavior. It represents a collection of individual electrons surrounded by clouds of charge that screen the electron fields within a distance of the order of magnitude of the screening length. This splitting of the density fluctuations corresponds to an effective separation of the Coulomb interaction into long-range and short-range parts, with the separation occurring at roughly the plasma screening length. Notably, the vast majority of atomic processes take place within the short-range region. Therefore, it is common to describe particle interactions using screening potentials, such as the Debye-Hückel potential [10-13] or atomic-state-dependent potential [14].

In the short-range region, compared with the energy level structures and photoionization processes [15-18], there have been relatively few studies on electron collision processes in plasmas. Early studies primarily relied on semi-empirical methods [19] or perturbative approaches such as the Born and distorted-wave approximations to investigate high-energy inelastic collisions [20-22]. For low-energy collisions involving resonance phenomena, existing close-coupling calculations in screened plasmas (R-matrix [23-25] and CCC [26,27] methods) primarily focus on neutral atomic targets. The key challenge lies in accurately extracting the short-range scattering phase shifts caused by the screened Coulomb potential of ions. Presently, the only attempt to obtain electron-ion collision data involves bypassing boundary condition issues by using a two-electron wave packet combined with the exterior complex scaling method [28]. However, this approach suffers from significant drawbacks, including high computational costs and applicability only to single-electron targets, limiting its extension to multi-electron systems. Thus, developing a comprehensive method to accurately calculate electron-ion collision processes in plasmas is critically important.

This study develops a comprehensive R-matrix close-coupling method for calculating electron-ion/atom collisions in plasma environments. We develop numerical methods to accurately extract scattering phase shifts from the asymptotic scattering wavefunctions. The reliability of the method is rigorously verified by comparing our results with both standard computational procedures [29] and published reference data [28]. As a specific application, we perform detailed calculations for electron collisions with hydrogen-like ions in Debye plasmas. The paper is organized as follows. Section II provides the theoretical framework, including the R-matrix method and our treatment of asymptotic scattering wavefunctions. Section III presents and analyzes our computational results. Finally, concluding remarks are given in Section IV.

## II. THEORETICAL METHOD

## A. Relativistic R-matrix method for screened interactions

The R-matrix method is a highly efficient close-coupling method used to investigate electron-ion/atom collision processes, as described in many references [30,31]. In short, the R-matrix scheme divides configuration space into two regions by a spherical boundary at $r_0$, known as the R-matrix boundary. In the inner region, $r < r_0$, the colliding electron is indistinguishable from the other $N$ target electrons. It is a many-electron problem, which is solved variationally as a whole to obtain the boundary matrix R(E) at $r_0$ for the (N+1)-electron system. The Dirac Hamiltonian for the (N+1)-electron system can be written in atomic units as,

$$H^{N+1} = \sum_{i=1}^{N+1}[-ic\alpha \cdot \nabla_i + (\beta-1)c^2 + V_{\text{nuc}}(r_i)] + \sum_{i=1}^{N}\sum_{j=i+1}^{N+1} V_{\text{ee}}(r_i, r_j). \quad (1)$$

In plasmas, $V_{\text{nuc}}(r_i)$ and $V_{\text{ee}}(r_i, r_j)$ are the screened Coulomb potentials. Taking the Debye-Hückel potential [10] as an example,

$$V_{\text{nuc}}(r_i) = -\frac{Z}{r_i}\exp\left(-\frac{r_i}{D}\right), \quad (2)$$

$$V_{\text{ee}}(r_i, r_j) = \frac{1}{|r_j - r_i|}\exp\left(-\frac{|r_j - r_i|}{D}\right)$$

$$= \frac{4\pi}{\sqrt{r_i r_j}}\sum_{l=0}^{\infty} I_{l+\frac{1}{2}}\left(\frac{r_<}{D}\right)K_{l+\frac{1}{2}}\left(\frac{r_>}{D}\right)$$

$$\times \sum_{m=-l}^{l} Y_{lm}^*(\theta,\phi)Y_{lm}(\theta,\phi), \quad (3)$$

where $D = \sqrt{k_B T_e/4\pi e^2 n_e}$ is the Debye screening length; $k_B$, $T_e$, and $n_e$ are the Boltzmann constant, electron temperature, and electron density, respectively. $I_{l+\frac{1}{2}}$ and $K_{l+\frac{1}{2}}$ are the modified Bessel functions of the first and second kind, respectively [32]; $r_< = \min(r_i, r_j)$, $r_> = \max(r_i, r_j)$, $Y_{lm}(\theta,\phi)$ are the spherical harmonics of rank $l$.

For the screened potentials in Eqs. (2) and (3), the angular coefficients for the one and two-electron integrals do not change relative to the pure Coulomb electron-nucleus and electron-electron interactions. Then we only need to modify the one and two-electron radial integrals by substituting the integrands with the corresponding radial part of the potentials shown in Eqs. (2) and (3), respectively, to obtain the inner region matrix elements and calculate the R-matrix R(E).

In the outer region, the exchange interactions between the colliding electron and the target electrons are negligible. The colliding electron mainly feels a local two-body potential with approximate long-range polarization potentials. When $r$ is large enough, relativistic effects on the wavefunctions become negligible, allowing the outer region to be described using nonrelativistic wavefunctions. In this case, the colliding electron can be described by ordinary differential equations [30],

$$\left[\frac{d^2}{dr^2} - \frac{l_i(l_i+1)}{r^2} - 2V(r) + k_i^2\right]u_i(r) = 2\sum_{\lambda=1}^{\lambda_{max}}\sum_{i'=1}^{n}\frac{C_{ii'}^{(\lambda)}}{r^{\lambda+1}}u_{i'}(r), \quad (4)$$

where $i$ and $i'$ are the channel indices, $u_i(r)$ is the radial wavefunction for channel $i$, $V(r)$ is the electron-nucleus interaction potential, $k_i$ is the wave number of the colliding electron, $C_{ii'}^{(\lambda)}/r^{\lambda+1}$ is multipole potential calculated by integrating the $V_{ee}(r_i, r_j)$ with the corresponding $i$, $i'$ channel wavefunctions, which are formed by coupling the $N$-electron target wavefunctions with the angular momentum and spin wavefunctions of the colliding electron. The coefficient $C_{ii'}^{(\lambda)}$ shown in Eq. (4),

$$C_{ii'}^{(\lambda)} = (2\lambda+1)r^{\lambda+\frac{1}{2}}K_{\lambda+\frac{1}{2}}\left(\frac{r}{D}\right)\left\langle\bar{\phi}_i\middle|\sum_{m=1}^{N}I_{\lambda+\frac{1}{2}}\left(\frac{r_m}{D}\right)/\sqrt{r_m}\times P_\lambda\cos\theta_{m,N+1}\middle|\bar{\phi}_j\right\rangle. \quad (5)$$

One can verify that when $D^{-1} = 0$,

$$C_{ii'}^{(\lambda)} = \langle\bar{\phi}_i|\sum_{m=1}^{N}r_m^\lambda\times P_\lambda(\cos\theta_{m,N+1})|\bar{\phi}_j\rangle, \quad (6)$$

which is the one used in previous unscreened calculations [33,34]. Note that unlike previous Ref. [24], which utilizes an approximate form of Eq. (5), i.e.,

$$C_{ii'}^{(\lambda)} = (2\lambda+1)r_0^{\lambda+\frac{1}{2}}K_{\lambda+\frac{1}{2}}\left(\frac{r_0}{D}\right)\left\langle\bar{\phi}_i\middle|\sum_{m=1}^{N}I_{\lambda+\frac{1}{2}}\left(\frac{r_m}{D}\right)/\sqrt{r_m}\times P_\lambda\cos\theta_{m,N+1}\middle|\bar{\phi}_j\right\rangle, \quad (7)$$

we directly employ Eq. (5) for calculations. Therefore, our treatment is more rigorous than previous approximate implementations under Debye screening.

For solving Eq. (4) with the channel-coupling terms on the right-hand side, one can use the well-developed asymptotic codes such as FARM [33] or STGF [34] codes for isolated neutral and ion systems. The final crucial step is the calculation of the scattering matrix by matching the asymptotic boundary condition with the R-matrix at the matching point $r_a$, i.e.,

$$F_{ij}(r_a) = \sum_{i'}R_{ii'}\left[r_a\frac{dF_{i'j}(r)}{dr} - bF_{i'j}(r)\right]_{r=r_a}, \quad (8)$$

with $F_{i'j}$ and $dF_{i'j}(r)/dr$ the channel radial wavefunction and its derivative, respectively. $b$ is a constant parameter in the R-matrix definition which usually takes zero. At large distances, where the potential is dominated by the centrifugal term, Eq. (4) has a zero right-hand side and can be written as,

$$\left[\frac{d^2}{dr^2} - \frac{l_i(l_i+1)}{r^2} - 2V(r) + k_i^2\right]u_i(r) = 0, \quad (4')$$

In this case, $F_{ij}$ can be expressed as,

$$F_{ij}(r) = S_i(kr)\delta_{ij} + C_i(kr)K_{ij}, \quad (9)$$

where $K_{ij}$ is the scattering $K$ matrix. $S_i(kr)$ and $C_i(kr)$ are the regular and irregular wavefunctions for Eq. (4'), respectively, which have the phase difference of $\pi/2$ and serve as a reference for the extraction of the short-range phase shift. For isolated neutral atoms, $S_i(kr)$ and $C_i(kr)$ are the Riccati-Bessel functions, whereas for ionic Coulomb scattering they become the corresponding Coulomb wavefunctions [29,35]; in both cases the asymptotic phase shifts are known analytically and implemented in standard R-matrix outer region programs [33,34]. However, for Debye-screened Coulomb potentials, the asymptotic wavefunctions are not available analytically. We therefore determine the screened asymptotic solutions by direct numerical calculation and match the channel wavefunctions at the R-matrix boundary to obtain the short-range scattering matrix.

## B. Numerical method for the asymptotic wavefunctions under the screened potentials

For a potential $V(r)$ that is less singular than $1/r^2$ at the origin, the behavior near $r = 0$ is dominated by the centrifugal term. We can expect two linearly independent solutions of Eq. (4'), $F_l$ and $G_l$, whose small-distance behavior is [36],

$$\begin{cases} F_l(r) \underset{r\to 0}{\propto} r^{l+1} \\ G_l(r) \underset{r\to 0}{\propto} r^{-l} \end{cases}. \tag{10}$$

Here $F_l$ is the physical, regular solution, while $G_l$ is an unphysical, irregular solution. In accordance with Eq. (10), the phase shift of the regular solution can be integrated outward from the origin, whereas the irregular solution diverges at the origin and cannot be integrated in the same manner. In the asymptotic region, the irregular solution is the linearly independent companion of the regular one and differs by a phase shift of $\pi/2$. Therefore, after propagating $F_l$ outward to a sufficiently large distance and determining the phase shift $\delta_l$, the asymptotic solutions ($F_l$ and $G_l$) can be constructed and integrated inward to any $r$ as needed.

For this purpose, we employ the logarithmic derivative method [37] to solve Eq. (4'). This method offers the advantages of simple programming, self-starting capability, and easy adjustment of the integration step size $h$ according to the energy. The truncation error between integration points scales as the fourth power of $h$. When the wavefunction is propagated from the origin to $r$, it can be expanded using Riccati-Bessel functions [38],

$$\begin{cases} F_l(r) = \alpha_l(r)[\cos\delta_l(r)\,S_l(kr) - \sin\delta_l(r)\,C_l(kr)] \\ F_l'(r) = k\alpha_l(r)[\cos\delta_l(r)\,S_l'(kr) - \sin\delta_l(r)\,C_l'(kr)]' \end{cases} \tag{11}$$

$$\alpha_l(r) = \exp\left(\frac{\gamma(r)}{2k}\right), \tag{12}$$

$$\gamma(r) = \int_0^r ds\, V(s)[S_l^2(ks) + C_l^2(ks)]\sin 2\left[-\tan^{-1}\frac{S_l(ks)}{C_l(ks)} + \delta_l(s)\right], \tag{13}$$

and the phase shift function is expressed as,

$$\delta_l(r) = \tan^{-1}\left[\frac{S_l(kr)L(r) - kS_l'(kr)}{C_l(kr)L(r) - kC_l'(kr)}\right]. \tag{14}$$

Here $L(r) = F_l'(r)/F_l(r)$ is the logarithmic derivative. The irregular solutions have the following asymptotic behavior,

$$\begin{cases} G_l \underset{r\to\infty}{=} \alpha_l(r)[\sin\delta_l(r)\, S_l(kr) + \cos\delta_l(r)\, C_l(kr)] \\ G_l'(r) \underset{r\to\infty}{=} k\alpha_l(r)[\sin\delta_l(r)\, S_l'(kr) + \cos\delta_l(r)\, C_l'(kr)] \end{cases}. \quad (15)$$

In practical calculations, it is not feasible to integrate to $r \to \infty$ to obtain asymptotic solutions. To improve computational efficiency, it is necessary to determine a finite maximum distance $r_{max}$ while ensuring the convergence of the Wronskian. By definition, Riccati-Bessel functions satisfy the Wronskian relation [39],

$$S_l(kr)C_l'(kr) - S_l'(kr)C_l(kr) = 1, \quad (16)$$

and the Wronskian relation between the regular and irregular wavefunctions is,

$$F_l(r)G_l'(r) - F_l'(r)G_l(r) = k\alpha_l^2(r). \quad (17)$$

Since the Wronskian of Eq. (4') with any potential $V(r)$ is a constant over all $r$. Then if $r$ is sufficiently large such that $\alpha_l^2(r)$ becomes $r$-independent, we regard this as the criterion for the convergence of $r_{max}$. Specifically, the difference

$$\Delta(r_{max}) = k[\alpha_l^2(r_{max} + h) - \alpha_l^2(r_{max})], \quad (18)$$

must be small enough to achieve our desired accuracy, typically $10^{-5}$.

Finally, we employ the Numerov method [40] to integrate the asymptotic wavefunctions inward and compute the wavefunction at the matching point $r_a$. At this point, we match the inner region wavefunctions and outer region solutions with the R-matrix to obtain the short-range scattering matrices.

Fig. 1 shows the Coulomb wavefunctions calculated with the present numerical method at $k^2$=5 a.u. for $Z$=1 and $l$=1. To ensure convergence, we set $\Delta(r_{max}) < 10^{-5}$ to determine the integration distance $r_{max}$. The wavefunction amplitude and phase shift show excellent agreement with results from the standard COULFG program [29]. After this validation, we apply the same procedure to the Debye potential. As shown in Fig. 1, for $D$=10 a.u., the Debye screening introduces a pronounced additional phase shift compared with the pure Coulomb case.

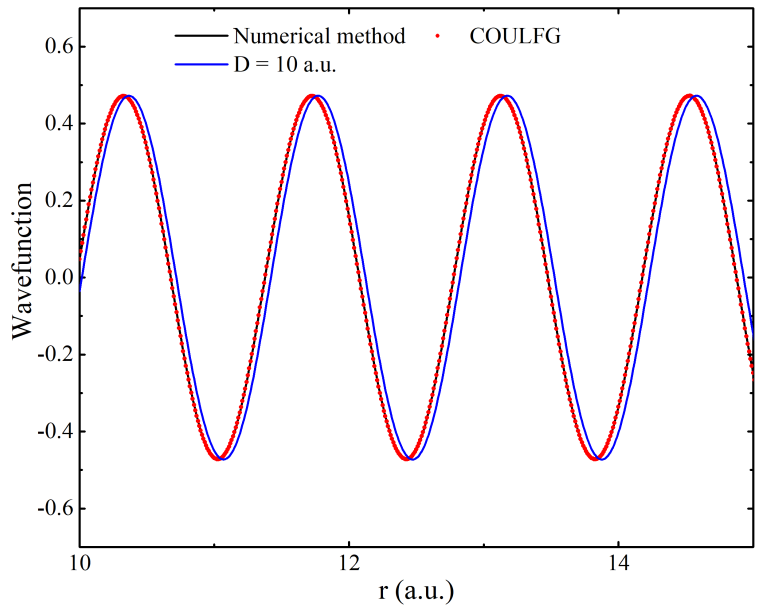

Fig. 1. Coulomb and Debye wavefunctions calculated using different methods.

## III. APPLICATION TO THE HYDROGEN-LIKE SYSTEM

### A. Collision strengths for He$^+$

For the variational calculation of the R-matrix inner region, the $1l$, $2l$, $3l$, and $4l$ ($l \leq 3$) orbitals are taken as spectroscopic orbitals, while the $5l$ and $6l$ ($l \leq 3$) orbitals serve as pseudo-orbitals. The hydrogen-like spectroscopic orbitals were first calculated, and then kept fixed in subsequent extended optimal-level calculations, in which the pseudo-orbitals were optimized sequentially with increasing principal quantum number to lower the $1s^2$ ground state energy of the electron-target complex, accounting for electron correlation effects. These orbitals are used to construct the $N$-electron target state wavefunctions. For the colliding electron, the continuum angular momentum satisfies $|\kappa| \leq 30$, and 60 continuum orbitals are employed for each $\kappa$-value to construct the ($N$ + 1)-electron collision system.

We employ the modified FARM code [33] to calculate the elastic and inelastic collision strengths from the ground state of He$^+$ in the low-energy region. In all calculations, the inner region R-matrix is calculated with a fixed screening length $D$=50 a.u. The outer region scattering wavefunctions under the screened interaction are constructed using three sets of asymptotic reference wavefunctions: unscreened Coulomb wavefunctions generated by the COULFG program [29], unscreened Riccati-Bessel wavefunctions, and Debye wavefunctions obtained from our numerical method.

Fig. 2(a) compares the inelastic collision strengths for the 1s $^2$S$_{1/2}$-2p $^2$P°$_{1/2}$ transition obtained using the three reference wavefunction schemes at different matching points. When the matching radius is chosen as $r_a$=42.51 a.u., minor discrepancies appear in the incident electron energy range 3.0-3.5 a.u. However, after propagating the R-matrix to a larger radius $r_a$=300 a.u., the results obtained with different asymptotic reference wavefunctions converge to nearly identical values. This behavior reflects the fact that inelastic scattering processes are dominated by short-range interactions occurring in the vicinity of the target. Once the matching radius extends beyond the interaction region, a screened short-range scattering matrix can be extracted that is essentially independent of the choice of asymptotic reference wavefunctions. Consequently, the extracted short-range scattering phase shifts become insensitive to the details of the asymptotic representation, explaining why previous calculations employing unscreened Coulomb wavefunctions were often able to produce reasonable inelastic collision strengths [20-22].

In contrast, Fig. 2(b) demonstrates that the elastic collision strengths for the 1s $^2$S$_{1/2}$-1s $^2$S$_{1/2}$ transition exhibit a pronounced dependence on the choice of asymptotic reference wavefunctions. Owing to the long-range nature of elastic scattering, the scattering phase shifts are strongly influenced by the asymptotic behavior of the wavefunctions. As a result, the elastic collision strengths obtained using Coulomb and plane-wave reference wavefunctions do not converge between the near and far matching points, although the resonance positions remain consistent. In the Coulomb results, a break in the vertical axis is introduced to illustrate the large discrepancies between calculations performed at different matching radii. By contrast, when Debye reference wavefunctions are employed, the extracted short-range scattering matrix becomes effectively independent of the matching radius, leading to stable and physically meaningful elastic collision strengths. These results demonstrate that an accurate treatment of the screened asymptotic behavior is essential for the reliable extraction of scattering phase shifts, particularly for elastic electron-ion collisions in plasma environments.

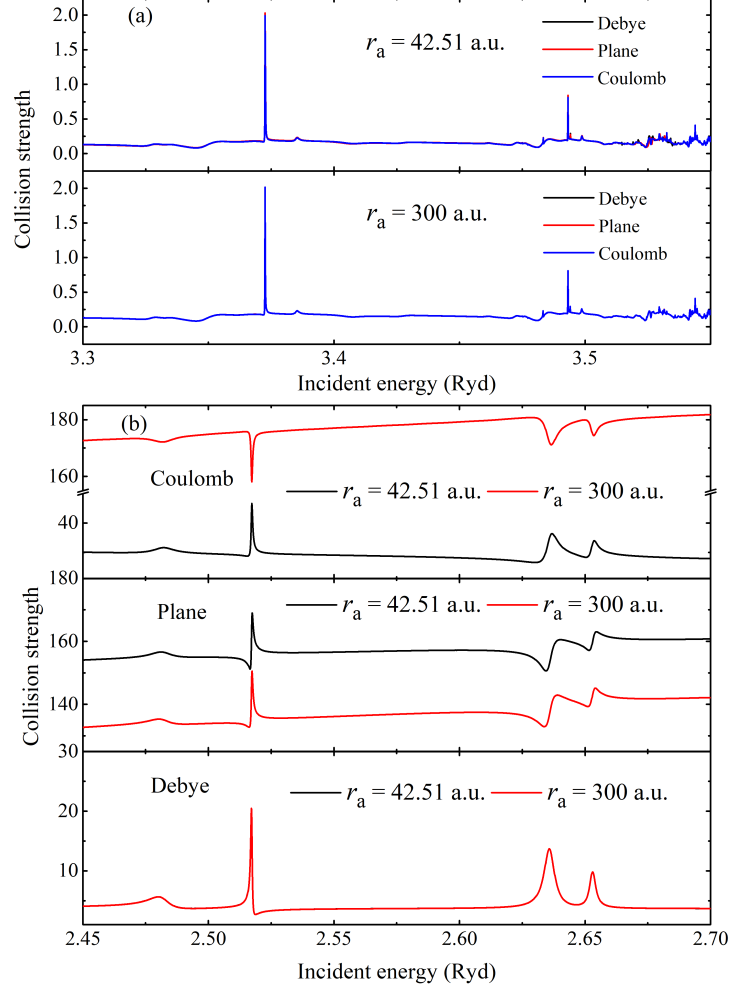

Fig. 2. Collision strengths for the 1s $^2S_{1/2}$-2p $^2P^o_{1/2}$ (a) and 1s $^2S_{1/2}$-1s $^2S_{1/2}$ (b) transitions of He$^+$, calculated using different reference wavefunctions (Coulomb, Debye, and plane-wave) at near and far R-matrix matching points.

Fig. 3 shows the ground state elastic collision strengths of He$^+$ for incident energies below the $n=2$ threshold, comparing our results with those obtained from the exterior complex scaling method [28]. Four dominant resonant structures can be observed near incident energies around 2.5 Ryd. Although the present calculations are relativistic while Ref. [28] is nonrelativistic, for He$^+$ the relativistic fine structure splittings are much smaller than the resonance widths, indicating a negligible relativistic effect for this low-energy electron collision process, and a direct comparison of the resonance structures is justified. These resonances are labeled as $_2(1,0)_2^{+1}S^e$, $_2(1,0)_2^{+3}P^o$, $_2(1,0)_2^{+1}D^e$, and $_2(0,1)_2^{+1}P^o$, corresponding to the 2s$^2$ $^1S^e$, 2s2p $^3P^o$, 2p$^2$ $^1D^e$, and 2s2p $^1P^o$ resonances, respectively. The resonance positions and background profiles at different screening lengths agree closely with Li's results [28].

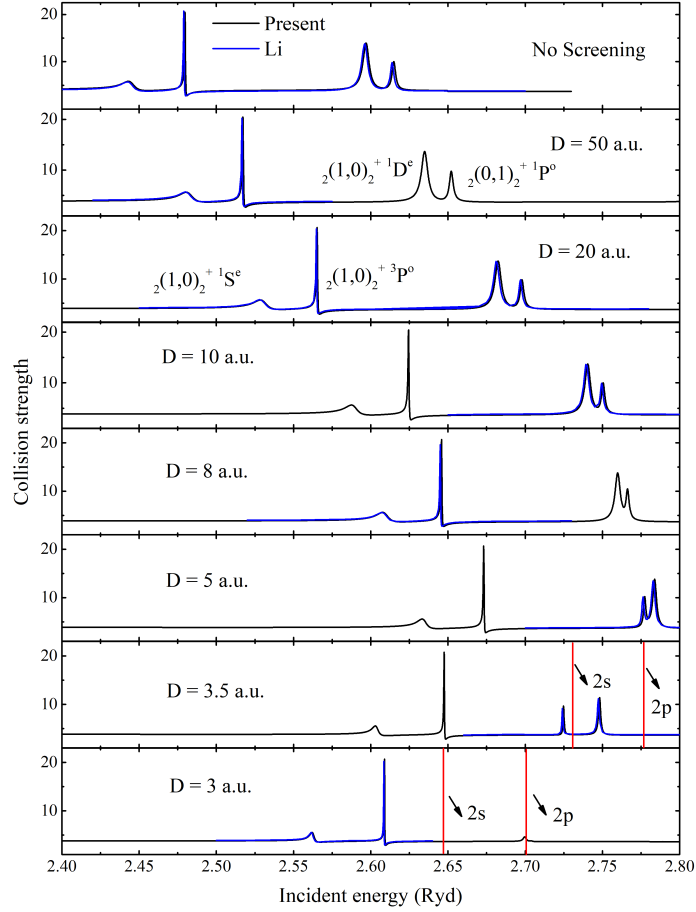

Fig. 3. Ground state elastic collision strengths of the He$^+$ ion for different screening lengths. Results from the reference are denoted as Li [28]. The red vertical lines indicate the excitation energies for the 2s and 2p states.

As the screening length decreases, the positions of the four resonances exhibit a systematic blue shift toward higher energies, followed by a red shift when the screening length is further reduced. At $D=5$ a.u., an inversion in the energy ordering of the $_2(1,0)_2^{+1}D^e$ and $_2(0,1)_2^{+1}P^o$ resonances is observed. Furthermore, when a resonance crosses the 2s threshold, its width and height change dramatically due to the transition from Feshbach to shape-type resonances. The $_2(1,0)_2^{+1}D^e$ resonance disappears quickly as the screening length decreases to $D=3$ a.u. These behaviors are consistent with the discussions in Refs. [23-25,28].

### B. Scaled effective collision strengths and electron-impact excitation rate coefficients for hydrogen-like ions

We also calculated inelastic and elastic collision strengths for excited states of other hydrogen-like ions in the energy region below the $n=3$ threshold. Fig. 4 shows the inelastic collision strengths for the 1s $^2S_{1/2}$-2p $^2P^o_{1/2}$ transition in C VI (left panel) and the elastic collision strengths for the 2s $^2S_{1/2}$ state in Ne X (right panel). Within the computational energy range, dense autoionization resonance structures associated with $3lnl'$ configurations are clearly identified.

Because the Debye potential decays faster than the Coulomb potential at large distances, bound state energies increase as the screening length $D$ decreases. The bound states will enter the continuum states when the screening length $D$ is less than the critical screening length, because only a finite number of bound states exist for a given screening length. As shown in Fig. 4, resonances associated with higher orbitals progressively disappear with decreasing $D$. Meanwhile, the excitation energy of the 3d target state decreases as the screening length is reduced, and this behavior becomes especially pronounced for $D \leq 30$ a.u. For example, at $D=3$ a.u., only the $3l3l'$ resonances remain in C VI.

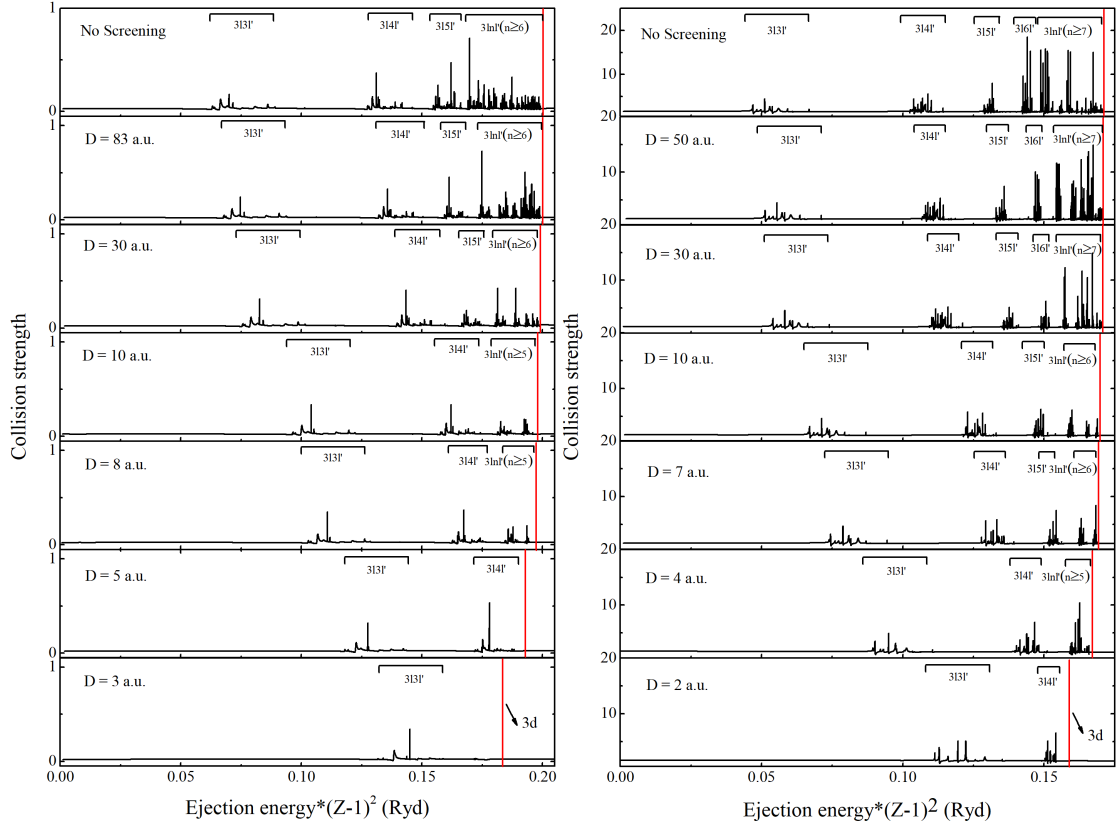

Fig. 4. The 1s $^2S_{1/2}$-2p $^2P^o_{1/2}$ inelastic collision strengths of the C VI ion (left panel) and the 2s $^2S_{1/2}$-2s $^2S_{1/2}$ elastic collision strengths of the Ne X ion (right panel) for different screening lengths. The red vertical lines indicate the excitation energy for the 3d state.

With decreasing screening length, the collision strengths are significantly modified, which directly affect electron-impact excitation rates. The effective collision strengths $\Upsilon_{ij}$ at electron temperature $T_e$ is obtained by integrating the collision strength over a wide energy range with a Maxwellian electron velocity distribution [41],

$$\Upsilon_{ij} = \int_0^\infty \Omega_{ij}\, e^{-\frac{\epsilon_j}{k_B T_e}} \mathrm{d}\left(\frac{\epsilon_j}{k_B T_e}\right), \qquad (19)$$

where $\Omega_{ij}$ is the collision strength for excitation from level $i$ to level $j$ and $\epsilon_j$ is the incident electron energy above the excitation threshold of the level $j$. The excitation rate coefficients are given by [42]:

$$q_{ij} = \frac{8.63*10^{-6}}{\sqrt{T_e}\, g_i}\, e^{-\frac{E_{ij}}{k_B T_e}} \Upsilon_{ij}, \qquad (20)$$

where $q_{ij}$ is the rate coefficient for electron-impact excitation, $E_{ij}$ is the energy difference between level $i$ and level $j$, and $g_i$ is the statistical weight of the initial state $i$.

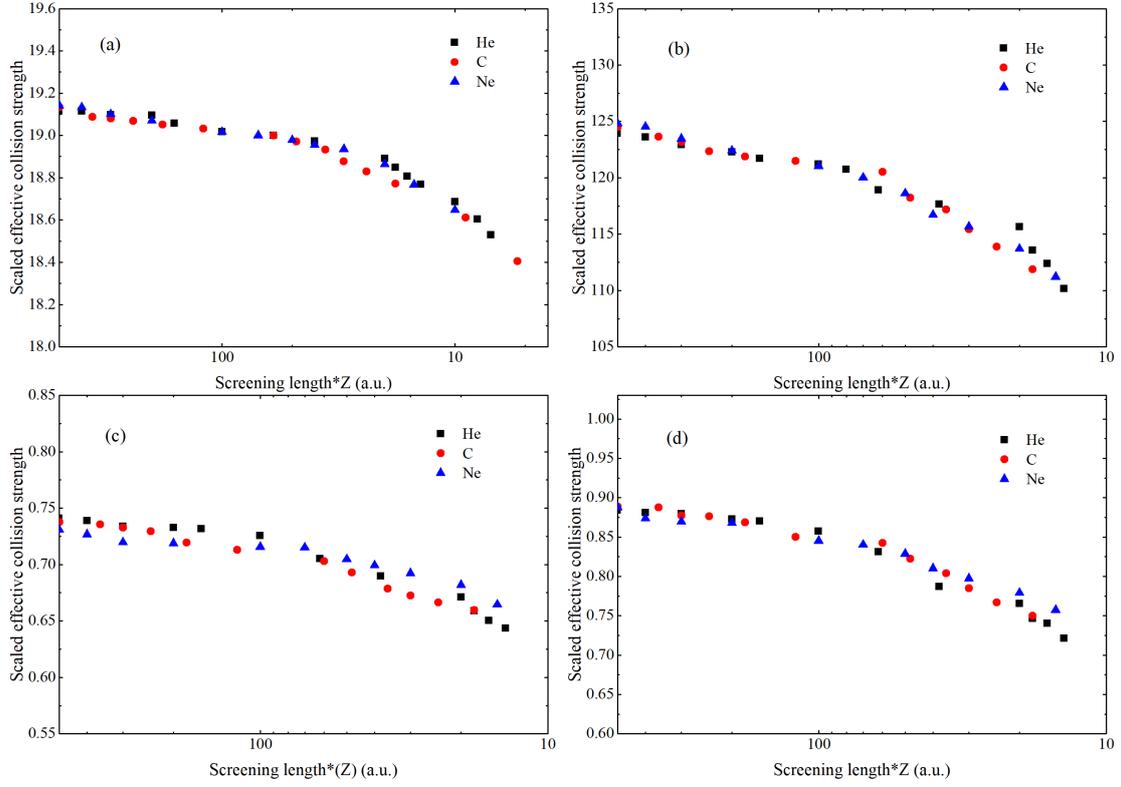

Fig. 5. Scaled effective collision strengths $\Upsilon * A(Z-1)^2 + B(Z-1) + C$ for four different transitions: (a) 1s $^2S_{1/2}$-1s $^2S_{1/2}$, (b) 2p $^2P^o_{1/2}$-2p $^2P^o_{1/2}$, (c) 1s $^2S_{1/2}$-2s $^2S_{1/2}$, and (d) 1s $^2S_{1/2}$-2p $^2P^o_{1/2}$.

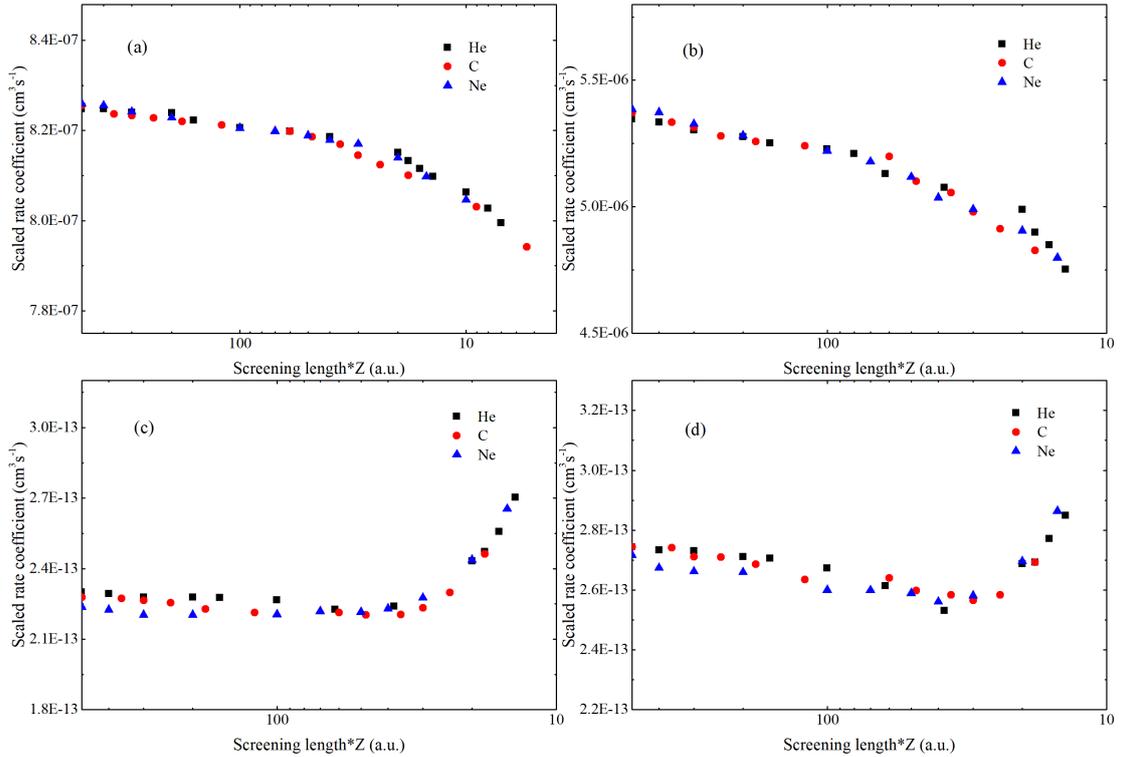

Fig. 6. Same as in Fig. 5 but for the scaled rate coefficients $q * Z(A(Z-1)^2 + B(Z-1) + C)$.

Fig. 5 shows the scaled effective collision strengths $\Upsilon * A(Z-1)^2 + B(Z-1) + C$ for four transitions as a function of the scaled plasma screening length $DZ$, calculated at a scaled temperature $T/Z^2 = 10^4$ K. The corresponding scaling coefficients A, B, and C for each transition are listed in Table 1. The overall trend is similar: the effective collision strengths decrease with decreasing screening length. For both elastic and inelastic processes, the scaled effective collision strengths decrease systematically as the screening length decreases, reflecting the progressive suppression of resonance structures at shorter $D$. Among the four cases, the ground state elastic scattering shows the weakest dependence on screening, consistent with the stronger binding of the ground state electron compared with excited states.

Table 1. Coefficients of the scaling law for different transitions.

| Transitions | Scaling coefficients | | |
| --- | --- | --- | --- |
| | A | B | C |
| 1s $^2$S$_{1/2}$-1s $^2$S$_{1/2}$ | 0.9208 | 2.3507 | 0.7043 |
| 2p $^2$P°$_{1/2}$-2p $^2$P°$_{1/2}$ | 0.9180 | 2.7245 | 1.4527 |
| 1s $^2$S$_{1/2}$-2s $^2$S$_{1/2}$ | 0.8324 | 3.3956 | 0.3853 |
| 1s $^2$S$_{1/2}$-2p $^2$P°$_{1/2}$ | 0.7691 | 2.1104 | 3.5284 |

Fig. 6 shows the scaled rate coefficients $q * Z(A(Z-1)^2 + B(Z-1) + C)$ for the same four transitions as a function of $DZ$ at $T/Z^2 = 10^4$ K. For elastic collisions, the rate coefficients follow trends similar to those of the effective collision strengths. For inelastic collisions, however, the scaled rate coefficients first decrease and then increase as the screening length decreases. This behavior stems from competing effects: at intermediate screening lengths, the weakening of resonance contributions reduces the rates, whereas for $D \leq 40$ a.u. the lowering of excitation thresholds allows more electrons to participate, causing the rates to rise sharply.

The scaled results exhibit a high degree of consistency across all three investigated ions, indicating that the present scaling may be used to estimate effective collision strengths and rate coefficients for other hydrogen-like ions under given plasma conditions through interpolation within the scaled curves, thereby facilitating their use in astrophysical and laboratory plasma research.

## IV. CONCLUSION

In this work, we develop a comprehensive relativistic R-matrix close-coupling method for electron-ion/atom collisions in plasma environments. Screened scattering wavefunctions in the asymptotic region are obtained by numerical methods and then matched to the inner region R-matrix to calculate the scattering matrices. The accuracy of the method is validated through two independent checks: (1) comparison of the calculated wavefunctions with standard reference programs, and (2) comparison of ground-state elastic collision strengths for He$^+$ with published results over a range of Debye screening lengths. Notably, the approach provides a self-consistent treatment of asymptotic matching for general screened interactions [14] and is readily extendable to multi-electron targets.

It is noteworthy that the applicability of the Debye screening for electron-electron interactions in atomic structure calculations has been debated [43,44], stemming from a spurious finding that a loosely bound H$^-$ ion between two electrons would remain bound even in the presence of a strong external plasma [17,45]. Our present theoretical framework offers a plausible explanation for this spurious result, suggesting that the Debye screening, when applied globally, can lead to unphysical outcomes. Since loosely bound or resonant H$^-$ states extend significantly beyond the reaction zone, the outer electron's identity is lost through interactions with plasma electrons, necessitating a collective many-body description. However, this does not invalidate the use of Debye screening within the reaction zone. Indeed, our results, as shown in Fig. 2, demonstrate the necessity of employing it to accurately obtain a physical short-range scattering matrix. Therefore, this work deepens our understanding of the appropriate application of Debye screening in electron-electron interactions.

For low-energy scattering of hydrogen-like ions in Debye plasmas, the collision strengths exhibit a strong dependence on the screening length, including systematic resonance shifts and progressive reductions of excitation thresholds. We find that the effective collision strengths decrease monotonically as screening becomes stronger, whereas inelastic excitation rate coefficients vary non-monotonically, decreasing first and then increasing at shorter screening lengths. Approximate unified scaling behaviors were observed in these statistical average quantities, which would significantly facilitate research in related plasma physics and astrophysics.

## ACKNOWLEDGMENTS

This work has been supported by the National Natural Science Foundation of China (Grant No. 12241410).